\documentclass[prl,twocolumn,amsmath,amssymb,nofootinbib,floatfix,superscriptaddress]{revtex4}

\usepackage{graphicx,bm}

\makeatletter
\def\graphicscale{\twocolumn@sw{0.3}{0.4}}
\def\graphicthreescale{\twocolumn@sw{0.3}{0.4}}

\begin{document}

\title{Phase diagram, symmetry breaking, and critical behavior \\
of three-dimensional lattice multiflavor scalar chromodynamics}

\author{Claudio Bonati} 
\affiliation{Dipartimento di Fisica dell'Universit\`a di Pisa 
       and INFN Largo Pontecorvo 3, I-56127 Pisa, Italy}

\author{Andrea Pelissetto}
\affiliation{Dipartimento di Fisica dell'Universit\`a di Roma Sapienza
        and INFN Sezione di Roma I, I-00185 Roma, Italy}

\author{Ettore Vicari} 
\affiliation{Dipartimento di Fisica dell'Universit\`a di Pisa
       and INFN Largo Pontecorvo 3, I-56127 Pisa, Italy}

\date{\today}

\begin{abstract}
We study the nature of the phase diagram of three-dimensional lattice
models in the presence of nonabelian gauge symmetries. In particular,
we consider a paradigmatic model for the Higgs mechanism, lattice
scalar chromodynamics with $N_f$ flavors, characterized by a
nonabelian SU($N_c$) gauge symmetry. For $N_f\ge 2$ (multiflavor case),
it presents two phases separated
by a transition line where a gauge-invariant order parameter
condenses, being associated with the breaking of the residual global
symmetry after gauging.  The nature of the phase transition line is
discussed within two field-theoretical approaches, the continuum
scalar chromodynamics and the Landau-Ginzburg- Wilson (LGW) $\Phi^4$
approach based on a gauge-invariant order parameter. Their predictions
are compared with simulation results for $N_f=2$, 3 and $N_c = 2$, 3,
and 4. The LGW approach turns out to provide the correct picture of
the critical behavior, unlike continuum scalar chromodynamics.
\end{abstract}

\maketitle

% ========================= BODY =========================

Local gauge symmetries are key features of theories describing
fundamental interactions~\cite{Weinberg-book} and emerging phenomena
in condensed matter physics~\cite{Sachdev-19}.  The large-scale
properties of three-dimensional (3D) gauge models and the nature of
their thermal or quantum transitions are of interest in several
physical contexts. For instance, they are relevant for
superconductivity~\cite{Anderson-63}, for topological order and
quantum transitions ~\cite{SBSVF-04,WNMXS-17,GASVW-18,SSST-19,GSF-19},
and also in high-energy physics, as they describe the
finite-temperature electroweak and strong-interaction transition which
occurred in the early universe~\cite{BVS-06} and which is presently
being investigated in heavy-ion collisions~\cite{STAR-05}.

We discuss a 3D model of interacting scalar fields with a nonabelian
gauge symmetry, which we may name scalar chromodynamics
or nonabelian Higgs model.  It provides a
paradigmatic example for the nonabelian Higgs mechanism, which is at
the basis of the Standard Model of the fundamental
interactions~\cite{SSBgauge}.  In condensed matter physics, 
it may be relevant for systems with emerging nonabelian 
gauge symmetries,
see, e.g., Ref.~\cite{GASVW-18}. It represents the natural extension of
 abelian Higgs models, which have been extensively studied
in various contexts, see, e.g.,
Refs.~\cite{Anderson-63,SBSVF-04,WNMXS-17,YHXV-18,YHVX-18,TS-18,PV-19}.
We will focus on the multiflavor case $N_f \ge 2$. For $N_f = 1$
scalar nonabelian models have 
have been carefully investigated, as they are relevant for
the finite-temperature behavior of the electroweak
theory~\cite{Nadkarni:1989na, Kajantie:1993ag, Buchmuller:1994qy,
  Kajantie:1996mn, Hart:1996ac}.  Much less is known about the phase
diagram and the nature of the 
transitions (symmetry-breaking pattern, universality class, etc.)
in the multiflavor case, and about the effective field
theory that describes the critical behavior.

In this paper we consider 3D lattice 
models of complex matricial scalar fields,
with $N_c\times N_f$ components, minimally coupled to an SU($N_c$)
gauge field~\cite{Wilson-74}. We investigate their phase diagram, for
various values of $N_f\ge 2$ and $N_c$, and the nature of their phase
transitions.  Our numerical results allow us to understand which
effective field theory provides the correct description of the phase
transition.  This study provides therefore information on the
field-theoretical approach to be used to analyze thermal and quantum
transitions in the presence of emergent nonabelian gauge symmetries.
Moreover, it may provide information on the finite-temperature phase
diagram of nonabelian gauge models involving scalar fields, that are
meant to describe the {\em new} physics beyond the Standard Model of
fundamental interactions.

Classical and quantum phase transitions have traditionally been
studied using statistical field theories~\cite{ZJ-book}.  Their
properties depend on the global symmetry of the model, the
symmetry-breaking pattern, and some other global properties, such as
the space dimensionality. In the presence of gauge symmetries, one may
think that also the gauge-symmetry group is a distinctive element that
should be specified to characterize the transition. As we shall
discuss, however, this is not necessarily true, as gauge modes are not
necessarily critical at the transition. For the same reason, at
variance with systems that only have global symmetries, there is not a
unique natural effective theory for the transition. One can, of
course, consider the continuum gauge theory that corresponds to the
lattice model. However, one may also consider a Landau-Ginzburg-Wilson
(LGW) $\Phi^4$ theory based on a gauge-invariant order-parameter field
with the global symmetry of the model~\cite{PW-84,PV-19}. Note that,
while in the first approach the gauge symmetry is still present in the
effective model, in the second one, gauge invariance does not play a
particular role beside fixing the order parameter. The LGW approach is
expected to be the correct one when the gauge interactions are
short-ranged at the transition. In the opposite case, instead, the
continuum gauge model should allow for the correct picture.  We recall
that the LGW approach was used to predict the nature of the
finite-temperature phase transition of hadronic matter in the massless
limit of quarks, implicitly assuming that the SU(3) gauge modes are
not critical~\cite{PW-84,PV-13}.  We compare the renormalization-group
(RG) predictions of the above-mentioned field-theory approaches with
numerical lattice results.  This study allows us to deepen our
understanding of their effectiveness and limitations, in particular
for the widely used LGW approach.

To investigate the above issues, we consider lattice scalar gauge
theories obtained by partially gauging a maximally symmetric model of
matrix variables $Z_{\bm x}^{af}$. We start from the lattice
action~\cite{footnote-unitlength}
\begin{eqnarray}
S_s = - J \sum_{{\bm x},\mu} {\rm Re} \,
{\rm Tr}\,Z_{\bm x}^\dagger Z_{{\bm x}+\hat\mu} \,,
\quad 
{\rm  Tr}\,Z_{\bm x}^\dagger Z_{\bm x} = 1\,, 
\label{ullimit}
\end{eqnarray} 
where $Z_{\bm x}^{af}$ are $N_c\times N_f$ complex matrices and the
sum is over all links of a cubic lattice ($\hat{\mu}$ are unit vectors
along the three lattice directions).  The model has a global O($N$)
symmetry with $N=2N_cN_f$. In particular, it is invariant under the
global SU($N_c$) transformations ${Z}_{\bm x} \to V {Z}_{\bm x}$,
$V\in$ SU($N_c$).  To make this symmetry a local one, we use the
Wilson approach~\cite{Wilson-74}.  We associate a SU($N_c$) matrix
$U_{{\bm x},\hat{\mu}}$ with each link, and consider the action
\begin{eqnarray}
&&S_g  =- \beta N_f \sum_{{\bm x},\mu} 
{\rm Re}\, {\rm Tr} \left[ Z_{\bm x}^\dagger \, U_{{\bm x},\hat{\mu}}
\, Z_{{\bm x}+\hat{\mu}}\right]  \label{hgauge}\\
&&\quad -
{\beta_g \over N_c} \sum_{{\bm x},\mu>\nu} {\rm Re} \, {\rm Tr}\,
\left[
U_{{\bm x},\hat{\mu}} \,U_{{\bm x}+\hat{\mu},\hat{\nu}} 
\,U_{{\bm x}+\hat{\nu},\hat{\mu}}^\dagger  
\,U_{{\bm x},\hat{\nu}}^\dagger\right]
\,,
\nonumber
\end{eqnarray}
where the second sum is over all lattice plaquettes.  Beside the
SU($N_c$) gauge invariance, the model has also a global U$(N_f)$
symmetry, ${Z}_{\bm x} \to Z_{\bm x} \, U$ with $U\in {\rm U}(N_f)$.
For $N_c = 2$ the global symmetry group is larger than
U($N_f)$. Indeed the action turns out to be invariant under the
unitary symplectic group Sp$(N_f) \supset \hbox{U}(N_f)$, see also
Refs.~\cite{Georgi-book,DP-14}. If one defines the 2$\times 2N_f$
matrix $\Gamma^{al}$
\begin{equation}
\Gamma^{af} =  Z^{af}\,, \qquad
\Gamma^{a(N_f + f)} =  \sum_{b} \epsilon^{ab} \overline{Z}^{bf}
\end{equation}
($\epsilon^{ab}=-\epsilon^{ba}$, $\epsilon^{12}=1$), one can show that
the action is invariant under ${\Gamma}^{al} \to \sum_{m} W^{lm}
\Gamma^{am}$ where $W\in$~Sp$(N_f)$, and $l,m=1,...,2N_f$.  For
$\beta_g=\infty$, the gauge fields are equal to the identity (modulo
gauge transformations), thus we recover the ungauged model in
Eq.~(\ref{ullimit}), i.e. the standard $N$-vector model with $N=2N_c
N_f$.

\begin{figure}[tbp]
\includegraphics*[scale=0.9]{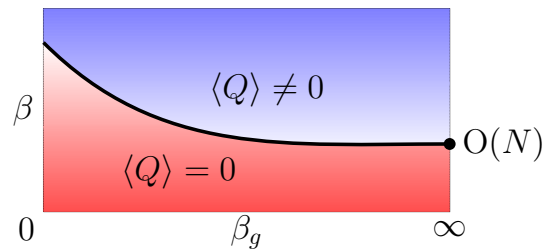}
\caption{Sketch of the phase diagram of 3D lattice scalar
  chromodynamics with $N_f\ge 2$.  The transition line is continuous
  for $N_f=2$ and of first order for $N_f\ge 3$.  We conjecture that
  its nature is the same for any finite $\beta_g$.  The endpoint for
  $\beta_g\to\infty$ is the O($N$) critical point ($N=2N_cN_f$).}
\label{phasediagram}
\end{figure}

For $N_f=1$ no transition \cite{footnote-Nf1} is expected for finite
$\beta_g$~\cite{OS-78,FS-79,DRS-80} (we have verified it numerically
up to $\beta_g = 6$), and long-range correlations should only develop
for $\beta_g\to\infty$ close to the O$(2N_c)$ critical point. For $N_f
\ge 2$ we find a transition line, that separates two different
phases---see Fig.~\ref{phasediagram}---which are characterized by the
behavior of gauge-invariant order parameter
\begin{equation}
Q_{{\bm x}}^{fg} = \sum_a \bar{Z}_{\bm x}^{af} Z_{\bm x}^{ag} -
{1\over N_f} \delta^{fg}\,,
\label{qdef}
\end{equation}
which is a hermitian and traceless $N_f\times N_f$ matrix.  The nature
of the transition depends on $N_f$ and $N_c$, while it does not depend
on the gauge coupling $\beta_g$.

Before presenting the numerical results, we discuss the predictions of
the statistical field theories which may describe the behavior along
the transition line.  We begin considering the continuum scalar
chromodynamics defined by the Lagrangian
\begin{eqnarray}
{\cal L} = {1\over 4 g^2} {\rm Tr}\,F_{\mu\nu}^2 + {\rm Tr} [(D_\mu
  Z)^\dagger (D_\mu Z)] + V({\rm Tr} Z^\dagger Z )\,,
\label{abhim}
\end{eqnarray}
where $V(X) = r X + {1\over 6} u X^2$, $F_{\mu\nu} = \partial_\mu
A_\nu - \partial_\nu A_\mu + [A_\mu, A_\nu]$ and $D_{\mu, ab} =
\partial_\mu\delta_{ab} + t_{ab}^c A_\mu^c$.  The RG flow in the space
of the renormalized couplings $u$ and $f\equiv g^2$ can be studied
perturbatively within the $\varepsilon\equiv 4-D$
expansion~\cite{WK-74}.  At one loop, the $\beta$ functions
read~\cite{GW-73,CEL-74,AY-94,Das-18,rescalingbetas}
\begin{eqnarray}
&&\beta_f(u,f) \equiv \mu {\partial f \over \partial \mu} = -
  \varepsilon f - (22N_c - N_f)\,f^2\,,
\label{betaf}\\
&&\beta_u(u,f) \equiv \mu {\partial u\over \partial \mu}
= - \varepsilon u + 
(N_f N_c +4)  \,u^2 \label{betau} \\
&&\quad  - {18 \,(N_c^2 -1)\over N_c} \, u f + 
{27 (N_c-1) (N_c^2 + 2 N_c-2)\over N_c^2}\,f^2\,.\qquad
\nonumber 
\end{eqnarray} 
A stable fixed point (FP) is only found for $N_f > N_{f}^*(N_c)$ with
$N_{f}^*(2) = 359 + O(\varepsilon)$ and $N_{f}^*(3) = 972.95 +
O(\varepsilon)$. Therefore, for any $N_f<N_f^*(N_c)$, and, in
particular, for small values of $N_f$, the transition is predicted to
be first order. We also note that for large $\beta_g$ the lattice
model (\ref{hgauge}) is expected to show significant crossover effects
due to the nearby O$(N$) transition point ($N=2N_cN_f$).  In the
theory (\ref{abhim}) such a crossover is controlled by the RG flow in
the vicinity of the O$(N)$ fixed point [$f^*=0$ and $u^*_N =
  \varepsilon/(N_f N_c +4)$], which is always unstable with respect to
the gauge perturbation~\cite{footnote-lambda}.

An alternative field-theoretical approach is provided by the LGW
framework~\cite{Landau-book,WK-74,Fisher-75,Ma-book,PV-02,PV-19}, in
which one assumes that the critical modes are associated with a
gauge-invariant composite operator. In the present case, the natural
order parameter is the operator $Q_{{\bm x}}^{fg}$ defined in
Eq.~(\ref{qdef}).  This is a nontrivial assumption, as it postulates
that gauge fields do not play a relevant role in the effective theory
of the critical modes.  The LGW fundamental field is correspondingly a
traceless hermitian matrix $\Psi^{fg}({\bm x})$, which can be formally
considered as the average of $Q_{\bm x}^{fg}$ over a large but finite
domain.  The LGW theory is obtained by considering the most general
4$^{\rm th}$-order polynomial consistent with the global symmetry:
\begin{eqnarray}
{\rm Tr}(\partial_\mu \Psi)^2 
+ r \,{\rm Tr} \,\Psi^2 
+  w \,{\rm tr} \,\Psi^3 
+  u\, ({\rm Tr} \,\Psi^2)^2  + v\, {\rm Tr}\, \Psi^4 
\label{hlg}
\end{eqnarray}
Continuous transitions may only occur if its RG flow has a stable FP.
For $N_f=2$, the cubic term vanishes and the two quartic terms are
equivalent, leading to the O(3)-symmetric LGW theory.  This implies
that the phase transition can be continuous, in the O(3) universality
class because of the mapping SO(3)$ = $SU(2)/${\mathbb Z}_2$. An
explicit O(3) order parameter is obtained by considering the real
vector variable $\varphi_{\bm x}^k = \sum_{fg} \sigma^k_{fg} Q_{\bm
  x}^{fg}$, where $\sigma^k$ are the Pauli matrices.  For $N_f\ge 3$,
the cubic $\Psi^3$ term is generically expected to be present.  For 3D
systems this is usually taken as an indication that phase transitions
are generically first order.~\cite{footnote-cubicterm}

The above arguments apply to generic $N_c>2$.  Since for $N_c=2$ the
global symmetry is Sp$(N_f)$, the order parameter is now a $2N_f\times
2N_f$ matrix given by
\begin{equation}
{\cal T}_{\bm x}^{lm} = 
\sum_a \overline{\Gamma}_{\bm x}^{al} \Gamma_{\bm x}^{am} -
     {\delta^{lm} \over 2 N_f} \sum_{an} 
     \overline{\Gamma}_{\bm x}^{an} \Gamma_{\bm x}^{an}\,,
\end{equation}
which can be expressed in terms of $Q_{\bm x}^{fg}$ defined in
Eq.~(\ref{qdef}) (${\cal T}_{\bm x}^{fg}=Q_{\bm x}^{fg}$ for
$f,g=1,...,N_f$) and of $D_{\bm x}^{fg} = \sum_{ab} \epsilon^{ab}
Z_{\bm x}^{af} Z_{\bm x}^{bg}$.  For $N_f=2$ the Sp(2) group is
isomorphic to the O(5) group.  The Sp(2) LGW theory is that of
O(5)-symmetric vector model.  Indeed, the order parameter can be
rewritten in terms of the three-component vector $\varphi_{\bm x}^k =
\sum_{fg} \sigma^k_{fg} Q_{\bm x}^{fg}$ and the real and imaginary
parts of the complex variable $\phi_{\bm x} = {1\over2} \sum_{fg}
\epsilon^{fg} D^{fg}_{\bm x}$. They form a five-component order
parameter. Thus we predict an O(5) critical behavior.  For larger
$N_f$, the Sp($N_f$) LGW theory contains cubic
interactions~\cite{BPV-20}, and therefore first-order transitions are
predicted.

Summarizing, the LGW approach based on a gauge-invariant order
parameter predicts that continuous transitions only occur for
$N_f=2$. They belong to the O(3) universality class for any $N_c\ge 3$
and to the O(5) universality class for $N_c=2$.  Instead, first-order
transitions are generically expected for $N_f\ge 3$ and any $N_c$.
Note that for $N_f=2$ the LGW predictions differ from those of the
continuum gauge theory (\ref{abhim}), as the latter predicts a
first-order transition (no FP for $N_f=2$).  The two theories
apparently also disagree for large values of $N_f$. The continuum
gauge theory admits the possibility of continuous transitions, since a
stable FP exists, while the LGW theory indicates first-order
transitions for any $N_f>2$.

In our numerical study~\cite{footnote-MC} we consider the model
(\ref{hgauge}) on a cubic lattice of size $L$ and periodic boundary
conditions.  We compute the correlation $G({\bm x}-{\bm y}) = \langle
{\rm Tr}\, Q_{\bm x} Q_{\bm y} \rangle$ of the composite operator
$Q_{\bm x}$ defined in Eq.~(\ref{qdef}), its susceptibility
$\chi=\sum_{\bm x} G({\bm x})$ and correlation length $\xi$,
\begin{eqnarray}
\xi^2 \equiv  {1\over 4 \sin^2 (\pi/L)}
{\widetilde{G}({\bm 0}) - \widetilde{G}({\bm p}_m)\over 
\widetilde{G}({\bm p}_m)},
\label{xidefpb}
\end{eqnarray}
where $\widetilde{G}({\bm p})=\sum_{{\bm x}} e^{i{\bm p}\cdot {\bm x}}
G({\bm x})$ and ${\bm p}_m = (2\pi/L,0,0)$.  We also consider the
Binder parameter
\begin{equation}
U = {\langle \mu_2^2\rangle \over \langle \mu_2 \rangle^2} , \quad
\mu_2 = {1\over L^6}  
\sum_{{\bm x},{\bm y}} {\rm Tr}\,Q_{\bm x} Q_{\bm y}\,.
\label{binderdef}
\end{equation}
At continuous transitions, RG invariant quantities, such as
$R_\xi\equiv\xi/L$ and $U$, behave as~\cite{ZJ-book,PV-02}
\begin{eqnarray}
R(\beta,L) = f_R(X) +  O(L^{-\omega})\,,\quad
X = (\beta-\beta_c)L^{1/\nu} \,,\quad
\label{scalbeh}
\end{eqnarray}
where $\nu$ is the correlation-length exponent, $f_R(X)$ is a
universal function (apart from a normalization of the argument), and
$\omega$ is the exponent associated with the leading scaling
corrections.  Moreover, since $R_\xi$ is a monotonic function,
Eq.~(\ref{scalbeh}) implies $U(\beta,L) \approx F_U(R_\xi)$, where
$F_U$ depends on the universality class only, without free
normalizations (once fixed the boundary conditions and the shape of
the lattice).

\begin{figure}[tbp]
\includegraphics*[scale=\graphicscale]{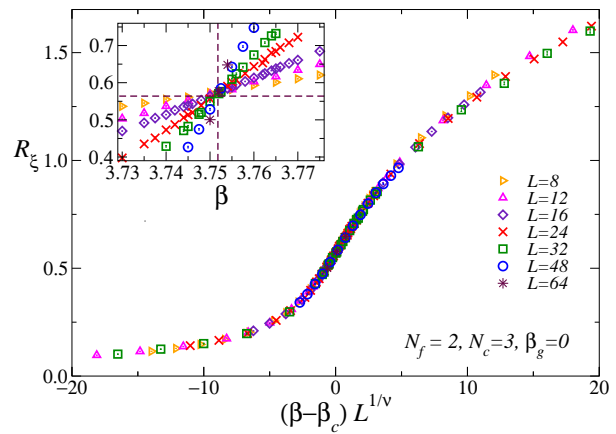}
\caption{ $R_\xi$ versus $(\beta-\beta_c)L^{1/\nu}$ for $N_f=2$,
  $N_c=3$, and $\beta_g=0$, up to $L=64$.  We use the O(3)
  value~\cite{HV-11,CHPRV-02} $\nu=0.7117(5)$ and
  $\beta_c=3.7518(2)$. The data collapse on a unique curve confirms
  the O(3) critical behavior. The inset reports $R_\xi$ versus
  $\beta$. The vertical dashed line corresponds to $\beta_c$, while
  the horizontal one corresponds to the critical value $R_\xi^*$,
  which is consistent with the O(3) value $R_\xi^*= 0.5639(2)$. }
\label{rxisca-f2c3}
\end{figure}

\begin{figure}[tbp]
\includegraphics*[scale=\graphicscale]{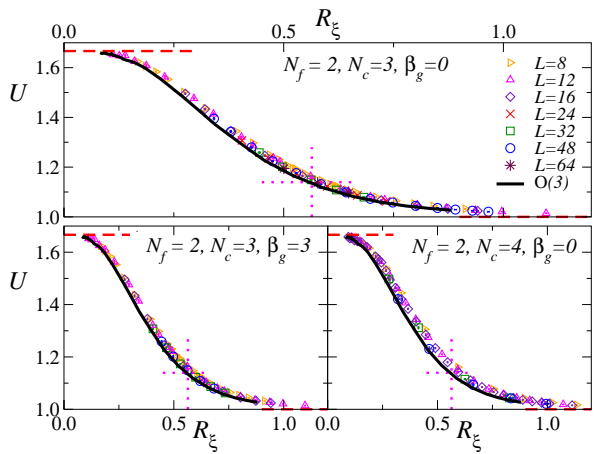}
\caption{Binder parameter $U$ versus $R_\xi$ for $N_f=2$ models:
  results for $N_c=3$, $\beta_g=0$ (top); for $N_c=3$, $\beta_g=3$
  (bottom left); for $N_c=4$, $\beta_g=0$ (bottom right).  In all
  cases the data appear to converge to the O(3) scaling curve,
  obtained by MC simulations of the O(3) vector
  model~\cite{BPV-20}. The dotted horizontal and vertical lines
  correspond to the O(3) critical values~\cite{CHPRV-02}
  $U^*=1.1394(3)$ and $R_\xi^*=0.5639(2)$.  }
\label{urxi-f2c34}
\end{figure}

The results for $N_f=2$, $N_c= 3$, and $\beta_g=0$ up to $L=64$, see
Fig.~\ref{rxisca-f2c3}, confirm that the transition at
$\beta_c=3.7518(2)$ is continuous, and belongs to the 3D O(3)
universality class, characterized by the universal
exponents~\cite{PV-02,HV-11,CHPRV-02,GZ-98} $\nu=0.7117(5)$, $\eta =
0.0378(3)$, $\omega=0.782(13)$.  We obtained analogous results for
$\beta_g=3$ [$\beta_c=3.203(1)$], and for $N_c=4$ at $\beta_g=0$
[$\beta_c=4.896(1)$], see Fig.~\ref{urxi-f2c34}, supporting the O(3)
nature of the transition in both cases. These results confirm the LGW
predictions. They lead us to conjecture that the phase diagram for
$N_f=2$ and $N_c\ge 3$ is characterized by a continuous transition
line, related to the condensation of the order parameter $Q_{\bm x}$,
which belongs to the O(3) universality class for any finite
$\beta_g$. For $\beta_g\to\infty$ the critical behavior turns into
that of O($N$) vector model with $N=4N_c$.

\begin{figure}[tbp]
\includegraphics*[scale=\graphicscale]{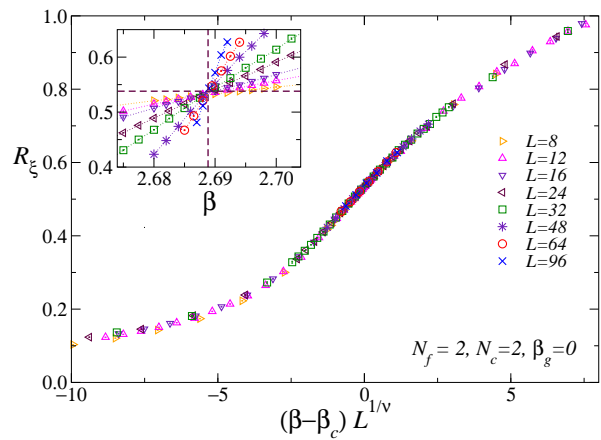}
\caption{ $R_\xi$ versus $(\beta-\beta_c)L^{1/\nu}$ for $N_f=2$,
  $N_c=2$, and $\beta_g=0$, up to $L=96$.  We use the O(5)
  value~\cite{HPV-05} $\nu=0.779(3)$ and $\beta_c=2.68885(5)$.  The
  data collapse supports the O(5) critical behavior.  The inset shows
  the data versus $\beta$: at $\beta_c$ (vertical dashed line) $R_\xi$
  is consistent with the O(5) value $R_\xi^*= 0.538(1)$ (horizontal
  dashed line).  Plots of $U$ versus $R_\xi$ as in
  Fig.~\ref{urxi-f2c34} also support the O(5) critical behavior. }
\label{rxisca-f2c2}
\end{figure}

We have also performed a FSS analysis for $N_f=N_c=2$ at $\beta_g=0$ (up
to $L=96$) and $\beta_g=2$ (up to $L=64$).  In both cases we observe
continuous transitions, at $\beta_c=2.68885(5)$ and
$\beta_c=1.767(1)$, respectively [note that $\beta_c=0.96339(1)$ in
  the O(8) vector model~\cite{DPV-15} obtained for
  $\beta_g\to\infty$]. Data are consistent with the O(5) universality
class, whose critical exponents
are~\cite{AS-95,HPV-05,FMSTV-05,CPV-03} $\nu=0.779(3)$,
$\eta=0.034(1)$, and $\omega = 0.79(2)$.  Numerical
results~\cite{BPV-20} for $\beta_g=0$ are shown in
Fig.~\ref{rxisca-f2c2}.  They confirm the predictions of the LGW
theory based on the enlarged global symmetry group Sp(2)$\approx$O(5).
We note that the enlarged O(5) symmetry may be seen as emerging from
the combination of an O(3) {\em magnetic}-like and of a U(1) {\em
  superfluid}-like order parameter~\cite{BPV-20}.  This provides a
possible mechanism for emergent O(5) symmetries, as also claimed in
various contexts, see, e.g.,
Refs.~\cite{GASVW-18,NSCOS-15,SPN-19,LJY-19,WSAMJ-14,XS-08,HKT-99,Henley-98}.

We finally mention that we also studied models for $N_f=3$, $N_c=2$
and 3, at $\beta_g=0$. The numerical results provide evidence of a
first-order transition in both cases~\cite{BPV-20}.  This is again
consistent with the predictions of the effective LGW theory
(\ref{hlg}).

In conclusion, we have investigated the phase diagram of the lattice
multiflavor scalar chromodynamics (\ref{hgauge}), for positive couplings $\beta$
and $\beta_g$.  This is a paradigmatic 3D model with a nonabelian
gauge symmetry.  For $N_f\ge 2$ the phase diagram is characterized by
two phases: a low-temperature phase in which the order parameter
$Q_{\bm x}^{fg}$ condenses, and a high-temperature disordered phase
where it vanishes.  Gauge and vector observables do not show
long-range correlations for any finite $\beta$ and $\beta_g$.  The two
phases are separated by a transition line driven by the condensation
of $Q_{\bm x}^{fg}$, as sketched in Fig.~\ref{phasediagram}, that ends
at the unstable O($N$) transition point with $N=2N_cN_f$ for $\beta_g
\to \infty$.  The gauge coupling $\beta_g$ does not play any
particular role: the nature of the transition is conjectured to be the
same for any $\beta_g$, as numerically checked for some values of
$\beta_g$.  Along the transition line only correlations of the
gauge-invariant operator $Q_{\bm x}^{ab}$ display long-range
order. Gauge modes are not critical and only represent a background
that gives rise to crossover effects.

The numerical results are compared with the predictions of the
continuum scalar gauge theory (\ref{hgauge}) and of the
gauge-invariant LGW theory (\ref{hlg}). They agree with those of the
LGW theory, showing that the LGW framework provides the correct
description of the large-scale behavior of these systems, predicting
first-order transitions for $N_f=3$, and continuous transitions for
$N_f=2$, which belong to the O(3) universality class for $N_c\ge 3$
and O(5) universality class for $N_c=2$.  On the other hand, the
results for $N_f =2$ are in contradiction with the predictions of the
continuum gauge model (\ref{hgauge}): as no stable FP exists for
$N_f=2$, one would expect a first-order transition.  There are at
least two possible explanations for this apparent failure.  A first
possibility is that it does not encode the relevant modes at the
transition.  A second possibility is that the perturbative treatment
around 4D does not provide the correct description of the 3D
behavior. The 3D FP may not be related to a 4D FP, and therefore it
escapes any perturbative analysis in powers of $\varepsilon$.  The
analysis of the behavior in the large-$N_f$ limit, where again the two
approaches give different results, may help to shed light on these
issues.  Similar issues for the multicomponent lattice scalar
electrodynamics are addressed in Ref.~\cite{PV-19,footnote-higgsU1}.

We have considered a paradigmatic lattice model obtained by gauging a
maximally symmetric scalar system. It would be interesting to consider
scalar theories with different global and local symmetries, and
different symmetry-breaking patterns. Their classification deserves
further investigation.

We mention that the LGW approach has been applied to the
finite-temperature transition of quantum chromodynamics (QCD) with
fermionic matter~\cite{PW-84,PV-13}.  The lattice-QCD numerical
results have only partially confirmed the LGW predictions, due to the
complexity of the simulations with fermions~\cite{Sharma-19}. Our
results support the effectiveness of the LGW approach, since the
derivation of the LGW theory is essentially independent of the bosonic
or fermionic nature of the matter fields.  We finally stress that the
results for lattice scalar chromodynamics may be particularly useful
in condensed matter physics, to understand the effects of emerging
nonabelian gauge fields~\cite{Sachdev-19,GASVW-18} at phase
transitions.

\hspace{0.5cm}

\emph{Acknowledgement}.
Numerical simulations have been performed on the CSN4 cluster of the
Scientific Computing Center at INFN-PISA.


\begin{thebibliography}{99}
\bibitem{footnote-unitlength} Note that the unit-length constraint of
  the $Z_{\bm x}$ variables can be considered as a particular limit of
  a quartic potential term $\sum_{\bm x} V( {\rm Tr} Z^\dagger_{\bm
    x}\,Z_{\bm x})$ with $V(X) = r X + {1\over 6} u\, X^2$.
\bibitem{footnote-Nf1} Fof $N_f=1$ the unit-length condition of the
  $Z$ variables, cf.~Eq.~(\ref{ullimit}), allows us to get rid of them
  by a gauge transformation, hence no transition is possible.
\bibitem{rescalingbetas} We rescale the couplings $u$ and $f$ as $u\to
  u/(24\pi^2)$ and $f\to f/(48\pi^2)$ to simplify the expressions.
\bibitem{footnote-lambda} The corresponding eigenvalue of the
  stability matrix is negative, $\lambda_f = - \varepsilon +
  O(\varepsilon^2)$ for any $N_f$ and $N_c$.
\bibitem{footnote-cubicterm} The presence of a cubic term in LGW
  theory is usually taken as an indication that the phase transitions
  occurring in the corresponding class of systems are first order.
  Indeed, a straightforward mean-field analysis shows that the
  transition is first order in four dimensions.  The nature of the
  transition should not change sufficiently close to four dimensions,
  as long as statistical fluctuations are small.  In particular, it is
  usually assumed that the 4D mean-field predictions also apply in
  three dimensions.
\bibitem{footnote-MC} We perform Monte Carlo simulations using an
  overrelaxed algorithm that consists in a mixture of microcanonical,
  heat-bath, and Metropolis updates.
\end{thebibliography}
\end{document}